# Sustainable Materials Discovery in the Era of Artificial Intelligence

Sajid Mannan[1], Rupert J. Myers[2], Rohit Batra[3], Rocío Mercado[4], Lothar Wondraczek[5,6], N. M. Anoop Krishnan[1,7,*]

[1]Department of Civil and Environmental Engineering, Indian Institute of Technology Delhi, Hauz Khas, New Delhi 110016, India
sajid.mannan@civil.iitd.ac.in
[2]Department of Civil and Environmental Engineering, Imperial College London, London, SW7 2AZ, UK
r.myers@imperial.ac.uk
[3]Department of Metallurgical and Materials Engineering, Indian Institute of Technology Madras, Chennai, 600036, India
rbatra@smail.iitm.ac.in
[4]Department of Computer Science and Engineering, Chalmers University of Technology & University of Gothenburg, Chalmersplatsen 4, 412 96 Gothenburg, Sweden
rocio.mercado@chalmers.se
[5]Otto Schott Institute of Materials Research, Lessingstrasse 14, 07743 Jena, University of Jena, Germany
[6]Center for Energy and Environmental Chemistry – CEEC, University of Jena, Philosophenweg 7, 07743 Jena, Germany
lothar.wondraczek@uni-jena.de
[7]Yardi School of Artificial Intelligence, Indian Institute of Technology Delhi, Hauz Khas, New Delhi 110016, India
[*]Corresponding author: krishnan@iitd.ac.in

**Abstract**

Artificial intelligence (AI) has transformed materials discovery, enabling rapid exploration of chemical space through generative models and surrogate screening. Yet current generative AI models for materials discovery, which now drive exploration of vast chemical and structural spaces, optimize candidates exclusively for structural stability and functional properties, with no integration of environmental assessment at any stage of the design loop. Prospective and ex-ante life cycle assessment methods exist and have been applied to emerging technologies, but they operate as standalone downstream analyses, not as active constraints within generative or active-learning pipelines. The result is that environmental feedback, even when produced, arrives after design decisions have been made rather than informing them. The disconnect between atomic-scale design and lifecycle assessment (LCA) reflects fundamental challenges: (i) data scarcity across heterogeneous sources, (ii) scale gaps from atoms to industrial systems, (iii) uncertainty in synthesis pathways, and (iv) the absence of frameworks that co-optimize performance with environmental impact. In this Perspective, we propose integrating upstream ML-assisted materials discovery with downstream LCA into the ML-LCA framework, comprising five components: information extraction for building materials-environment knowledge bases, harmonized databases linking properties to sustainability metrics, multi-scale models bridging atomic properties to lifecycle impacts, ensemble prediction of manufacturing pathways with uncertainty quantification, and uncertainty-aware optimization enabling simultaneous performance-sustainability navigation. Case studies spanning polymers, glass, photoresists, and cement demonstrate both necessity and feasibility while identifying material-specific integration challenges. Realizing ML-LCA requires coordinated advances in data infrastructure, ex-ante assessment methodology, multi-scale modeling, and uncertainty-aware optimization, enabling discovery of materials that are sustainable by design.

## 1. Introduction

Recent progress in artificial intelligence (AI) and machine learning (ML) is transforming how materials are designed, evaluated, and ultimately deployed[1–3]. AI-enabled workflows integrating high-throughput density functional theory (DFT)[4], machine-learned interatomic potentials (MLIPs)[5,6], and surrogate models now probe chemical and structural spaces that were computationally or experimentally inaccessible just a decade ago. These approaches have demonstrated notable success in domains such as battery materials, catalysis, and electronic applications, where data-driven screening routinely identifies promising candidates for targeted synthesis[7–9]. Despite these advances, the AI-driven materials discovery paradigm remains predominantly performance-centric[10], prioritizing metrics such as stability, novelty, and material properties, while aspects such as performance at the product scale or broader societal and environmental implications are rarely incorporated into early-stage decision-making.

This performance-first approach stands in stark contrast to the more mature frameworks developed in adjacent fields. Drug discovery, for instance, operates within a well-defined, closed-loop workflow where molecular candidates progress through distinct stages, from target identification and lead optimization to preclinical testing and clinical trials, ultimately culminating in regulatory approval (Figure 1(a))[11,12]. More importantly, considerations of safety, toxicity, bioavailability, and efficacy are integrated at every stage, not as afterthoughts, but as integral design constraints[13]. Materials discovery lacks such a coherent framework (Figure 1(b)) and the workflow remains fragmented. Computational screening identifies candidates with desired properties, experimental synthesis validates feasibility, and characterization confirms performance; however, the pathway from laboratory-scale demonstration to industrial deployment and real-world application remains poorly defined. Most of the research on AI-enabled materials discovery remain focused on generative modeling of crystals[14], and a few have synthesized those predicted materials[15,16]. This covers only the first two stages of materials discovery (see Figure 1(b)). Unlike drugs, which have clearly defined targets (biological pathways) and success metrics (clinical endpoints), materials serve diverse functions across vastly different application contexts. Further, socio-economic, environmental, and supply chain considerations in materials discovery are treated as auxiliary, post hoc analyses rather than integral components of the design objective; an issue that becomes increasingly problematic as AI accelerates the pace of materials innovation faster than sustainability considerations are incorporated (see Supplementary Section 2 for details).

Sustainability is a multidimensional concept, encompassing environmental, social, and economic considerations. While environmental life cycle assessment (LCA) strictly measures the physical and ecological damage, life cycle sustainability assessment (LCSA) with supply risk expands that scope to evaluate financial costs, social impacts, and geopolitical supply chain vulnerabilities simultaneously. In the context of design and discovery of new materials, these dimensions may be quantifiable through different metrics contingent upon the context. Here, we primarily focus on environmental metrics through LCA, which can quantify environmental impacts across a material's full lifecycle (Figure 1(c)); from raw material extraction through manufacturing, use, and end-of-life management[17]. By tracking resource consumption, greenhouse gas emissions, toxicity, and ecosystem impact at each stage, LCA enables holistic sustainability evaluation. Note that other metrics such as the Herfindahl-Hirschman Index (HHI)[18,19] quantifying supply chain risk, while valuable for identifying geopolitical vulnerabilities, operate outside standard LCA frameworks. Prospective and ex-ante LCA methods, which estimate lifecycle impacts from laboratory or pilot data using proxy technologies, process modeling, and scaling relationships, have been applied to emerging materials and chemical technologies[20–29]. However, they remain completely absent in the generative and active-learning pipelines that now define the frontier of AI-driven materials discovery. Models such as MatterGen[30] and GNoME[31] generate and screen candidates at scales of millions of structures, optimizing for stability, symmetry, and target properties; environmental constraints play no role in the design objective, the training data, or the candidate filtering. Integrating LCA into these pipelines, as a continuously updated design constraint rather than a downstream evaluation, requires automating the most labor-intensive steps of prospective LCA through machine learning, which is the specific contribution of the ML-LCA framework proposed here (see Supplementary Section 3 for more details).

Integrating LCA into early-stage discovery would transform environmental impact from post-hoc evaluation to active design constraint (Figure 1(c)). However, achieving this integration is non-trivial. LCA operates at macroscopic scales relating to industrial processes and product systems, while materials discovery operates at atomic and molecular scales. The data mismatch is severe: materials databases contain millions of computed structures; LCA databases document thousands of industrial processes. For novel materials, LCA must predict not only properties but also synthesis routes, manufacturing pathways, and in-use behavior—information inherently uncertain for materials that do not yet exist.

In this Perspective, we examine the major obstacles to integrating sustainability into AI-driven materials discovery and outline a roadmap toward ML-LCA frameworks that can co-optimize functional performance and sustainability (Figure 1c). Although environmental LCA serves as the main sustainability metric throughout, the proposed ML-LCA framework is in principle extendable to other quantitative indicators such as supply chain risk (e.g., the Herfindahl-Hirschman Index) and techno-economic metrics. Throughout this Perspective, we distinguish three modes of LCA in the context of materials discovery: (i) retrospective LCA, applied to established materials with measured industrial data; (ii) prospective or ex-ante LCA, applied to emerging materials using estimated inventories and scaling relationships; and (iii) closed-loop ML-LCA integration, in which automated prospective LCA operates within the generative discovery pipeline as an active design constraint. The ML-LCA framework addresses the third mode, building on the methodological foundations of the second. We analyze the data challenges, scale mismatches, and uncertainty management requirements inherent to this integration. Through case studies spanning polymers, glass, photoresists, and cement, we demonstrate both the necessity and feasibility of this approach. Our goal is not to propose a complete solution but to chart a course toward discovery workflows where sustainability is intrinsic rather than incidental. A Glossary of terms drawn from materials science, AI/ML, and LCA communities is provided in the Supplementary Section 6 to support readers across disciplines.

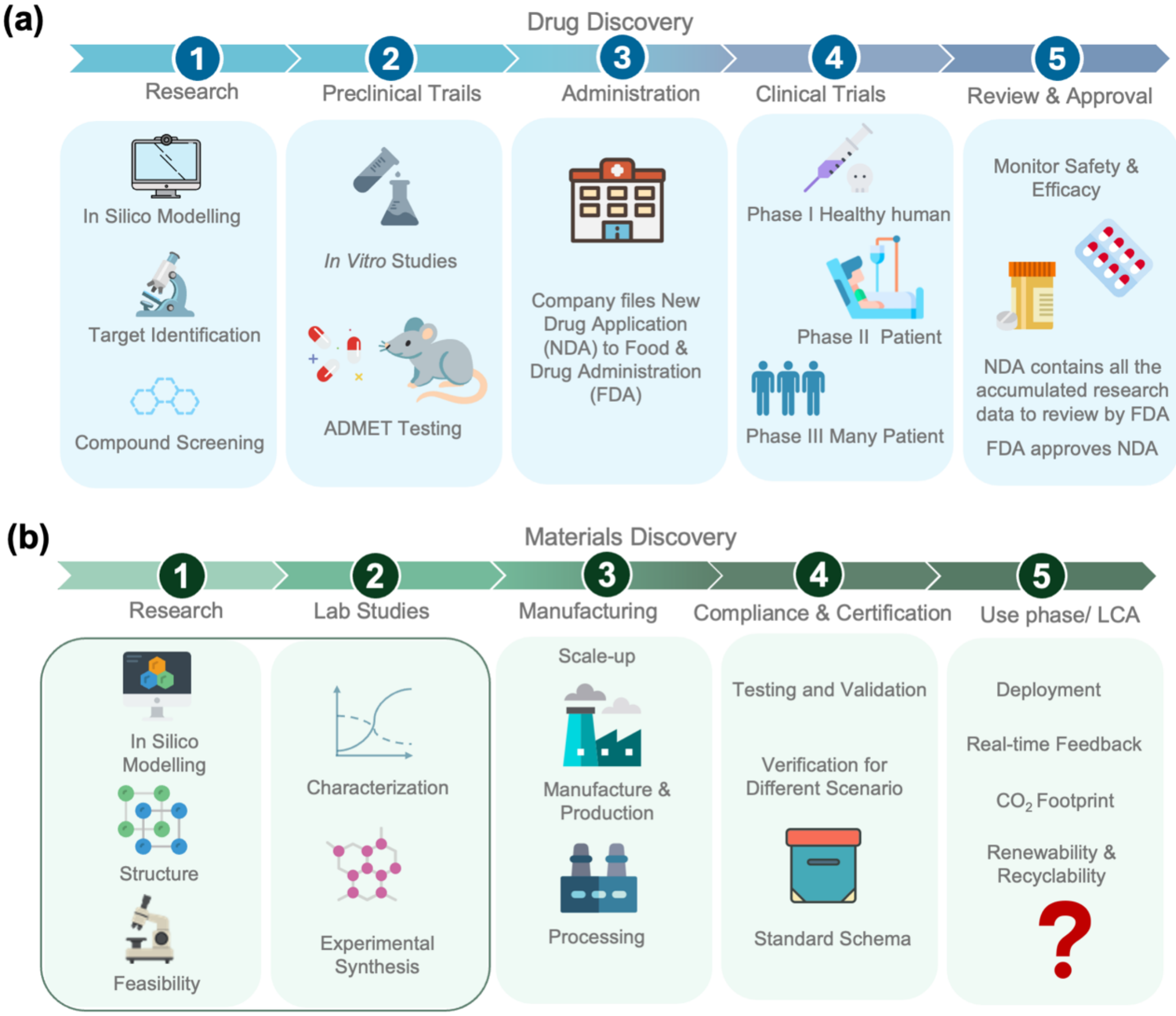

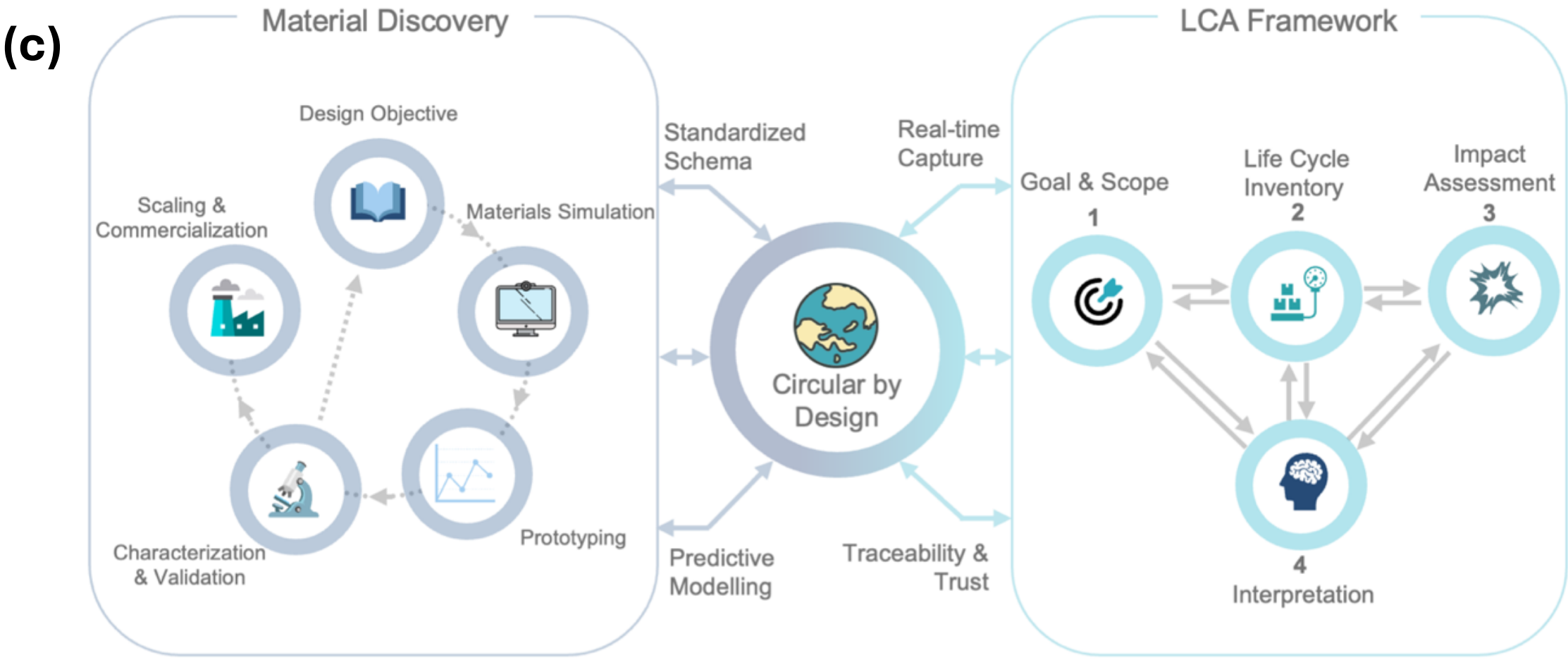

**Figure 1. Contrasting discovery paradigms**. (a) Drug discovery follows a well-defined workflow from molecular design through regulatory approval, with safety, toxicity, and efficacy integrated at each stage. (b) Materials discovery remains fragmented and mainly focused on the first two blocks (highlighted within the box), with a poorly defined pathway from computational screening to real-world deployment. Materials impact, determined not only by intrinsic properties but by their role in complex life cycles and systems, is rarely available during early-stage discovery. (c) (Left) Materials discovery currently follows a cyclical workflow from design objectives through simulation, prototyping, characterization, and commercialization, operating largely independently of sustainability assessment. (Right) Life cycle assessment follows its four-stage framework (goal and scope, inventory, impact assessment, interpretation) but typically engages only after materials reach commercial scale. (Center) The circular-by-design paradigm envisions systematic integration through standardized data schemas enabling seamless information exchange, real-time capture of lifecycle implications to provide early-stage feedback, predictive modeling that bridges scale gaps from atoms to industrial systems, and traceability mechanisms that maintain transparency about assumptions and uncertainties. Current practice shows minimal connection between these workflows (thin arrows); the proposed integration requires robust bidirectional information flow (thick arrows) where materials discovery informs LCA projections and LCA insights guide design decisions.

## 2. Challenges in sustainable materials discovery

Materials discovery and sustainability assessment operate in parallel, rarely intersecting until late-stage deployment (Figure 1(c)). Closing this gap requires a “circular-by-design” approach (Figure 1(c)), which embeds LCA as an active constraint throughout discovery workflow rather than appending retrospectively (see Supplementary Sections 1 and 4 for more details). Realizing this is hindered by several challenges outlined below.

### Data modality and multi-fidelity integration

Integrating environmental LCA into materials discovery requires workflows capable of handling multi-fidelity data[32]—information drawn from sources that differ in accuracy, resolution, and cost, high-fidelity DFT calculations alongside lower-fidelity empirical simulations or engineering estimates—from heterogeneous sources. Materials databases such as the Materials Project[33] contain millions of computed crystal structures with properties predicted from first-principles calculations, while comprehensive LCA databases[34,35] (ecoinvent, GaBi), document lifecycle inventories for only thousands of industrial processes, primarily mature technologies with decades of production history[36]. More challenging still is the heterogeneity of data types that must be integrated. Manufacturing process data from pilot plants with uncertain scale-up trajectories must be reconciled with long-term degradation predictions from accelerated aging tests. Each data type has different reliability, temporal validity, and uncertainty characteristics, and no existing framework integrates them systematically within an LCA pipeline.

### Data scarcity, consistency, and completeness

High-quality synthesis and property data for materials are often missing from databases[37], but the challenge extends beyond quantity to fundamental availability leading to promising approaches such as ex-ante LCA[20,23,38]. The key challenge in ex-ante LCA, the prospective assessment of technologies before industrial-scale data is available, is obtaining a reliable description of the product system from laboratory or pilot-scale demonstrations alone. Ex-ante LCA studies are defined[20] as those that: (i) scale-up an emerging technology using likely scenarios (e.g. using expert help, extreme views, learning curves for similar technologies) of future performance at full operational scale, or (ii) compare the emerged technology at scale with the evolved incumbent technology. Upscaling methods applicable at this stage[20,23] include parameter extrapolation, technology learning curves, and process-based projections. Several definitions used for sustainability-relevant data are inconsistent[20,39], making the compilation of datasets challenging and hindering ML applications that require standardized inputs. For example, a consistent definition of biodegradable polymers (or materials) is lacking. Biodegradability is reported in different ways: weight loss, reduction in strength, carbon mineralization, and visual disintegration. Similar issues exist for dissolution rates, ecotoxicity endpoints, and recyclability metrics. Without standardized definitions and measurement protocols, integrating data from different sources becomes unreliable, and ML models trained on inconsistent data yield unreliable predictions.

### Scale gaps from atoms to industrial systems to in-use performance

A fundamental obstacle is the limited understanding of environmental footprint when scaling up from molecules or crystals to industrial-scale production and ultimately to real-world structures in use. Furthermore, little knowledge exists about the in-use stages of newly discovered materials—durability, degradation pathways, maintenance requirements, end-of-life behavior—yet these factors determine lifecycle impacts. Ex-ante LCA methodology currently focuses on the effects of technology upscaling or future implementation but struggles to estimate unit process types for truly novel materials. Materials with similar bulk composition may be processed in very different ways, for example, graphite via the Acheson process (a high-temperature electric furnace) versus graphene via chemical vapor deposition, each with orders-of-magnitude differences in energy requirements and emissions profiles.

The LCA model needs to rely on the parameter combinations that the ML model has attributed to a material. For example, if the ML model predicts that heating a certain mixture of solid feedstocks to 1100°C and then cooling at a rate of 1°C per second yields a certain phase, then the LCA model should include mixing, heating, and cooling processes with types appropriate for solid materials handling. A search can be conducted to find a comparable product system for an established material—for this example, silicate glass production—which can then be used as the proxy technology to estimate its industrial-scale unit process data or life cycle inventory. Therefore, there should be close integration of the AI and LCA models, where the output of the AI models directly informs LCA process selection and parameterization.

**Synthesis pathway prediction**

Retrosynthesis[16]—predicting viable synthesis routes from target materials to available precursors—represents a major challenge that has been addressed by several models in organic chemistry and drug discovery but needs to be extended to broader classes of materials. Materials synthesis is far more diverse than organic synthesis: solid-state reactions, chemical vapor deposition, sol-gel processing, hydrothermal synthesis, mechanochemical activation, and many other methods each operate under different conditions with different energy requirements, waste streams, and environmental profiles. Moreover, retrosynthesis predictions must include durability considerations and in-use performance. A synthesis route that produces a material with excellent initial properties, but poor long-term stability may generate a higher lifecycle burden through frequent replacement than an alternative route that produces a more durable material. Yet predicting long-term degradation from atomistic calculations or short-term experiments remains highly uncertain, particularly for novel chemistries lacking empirical degradation databases or validated accelerated aging protocols.

**Cross-domain reward functions and multi-objective optimization**

A major challenge toward incorporating sustainability into materials discovery, through reward function or optimization objective, is in the construction of meaningful multi-objective functions across domains. For example, in the case of autonomous experiments, a poorly chosen reward function at any intermediate step may inadvertently drive whole experiments towards misaligned outcomes—a phenomenon known as reward hacking. Misalignment often arises when long-term goals such as discovery or sustainability are reduced to short-term targets[40]. Objectives have incommensurable units, for example, minimizing carbon footprint (kg $CO_2$-eq), minimizing embodied energy (MJ), maximizing recyclability (dimensionless), and different stakeholders weigh these objectives differently. Objectives also have vastly different uncertainties: DFT-predicted properties may have well-characterized error bounds, while lifecycle impacts for novel materials involve compounded uncertainties from synthesis route prediction, scale-up extrapolation, and future background system projections. Furthermore, sustainability objectives often exhibit non-linearities and threshold effects. A material with 95% recyclability may have vastly different end-of-life impacts than one with 60% recyclability if the threshold for viable recycling infrastructure is 90%. Capturing these nonlinearities in differentiable optimization frameworks suitable for gradient-based ML training presents technical challenges that current single-objective or simple Pareto-frontier approaches do not adequately address. Prior works, at the intersection of LCA and mathematical programming, demonstrate that these challenges are tractable for defined process systems. Earlier approaches employed mixed integer programming[41] to minimize a set of LCA impacts, while later works extended this to multi-objective optimization integrated hybrid LCA minimizing, for instance, greenhouse gas emissions[42]. These frameworks address well-characterized process spaces; extending them to the vast, partially defined chemical spaces that generative ML models explore requires approaches that can handle uncertain process mappings alongside uncertain property predictions.

**Uncertainty propagation across prediction stages**

Uncertainty is inherent in prospective assessment, but the compounding of uncertainties across multiple prediction stages creates challenges for ML-LCA[43]. For instance, generative models introduce uncertainty in stability and synthesizability; surrogate and retrosynthesis models add uncertainty from limited training data and route feasibility; and process and LCA models compound these through uncertainty in inventories[44], scale-up extrapolation and future background system projections. Uncertainty sources differ in character. Epistemic uncertainty can in principle be reduced through better models and data; aleatoric uncertainty (e.g., synthesis yield variability) cannot be reduced but can be characterized; and scenario uncertainty reflects alternative plausible futures. The canonical typology[45] used for characterizing and propagating these uncertainty types in LCA, and their framework, including Monte Carlo simulation, analytical error propagation, and fuzzy set methods, remains the reference starting point. This framework was further extended[46,47] by establishing statistical conventions for interpreting probabilistic results in comparative assessments, leading to comparative probabilistic LCA[47], directly relevant to the multi-material screening that ML-LCA requires. Current approaches in materials science and LCA address these separately, whereas ML-LCA demands their joint propagation through the full multi-scale, multi-model chain—a task that has not been systematically addressed.

## 3. The ML-LCA framework: architecture and implementation roadmap

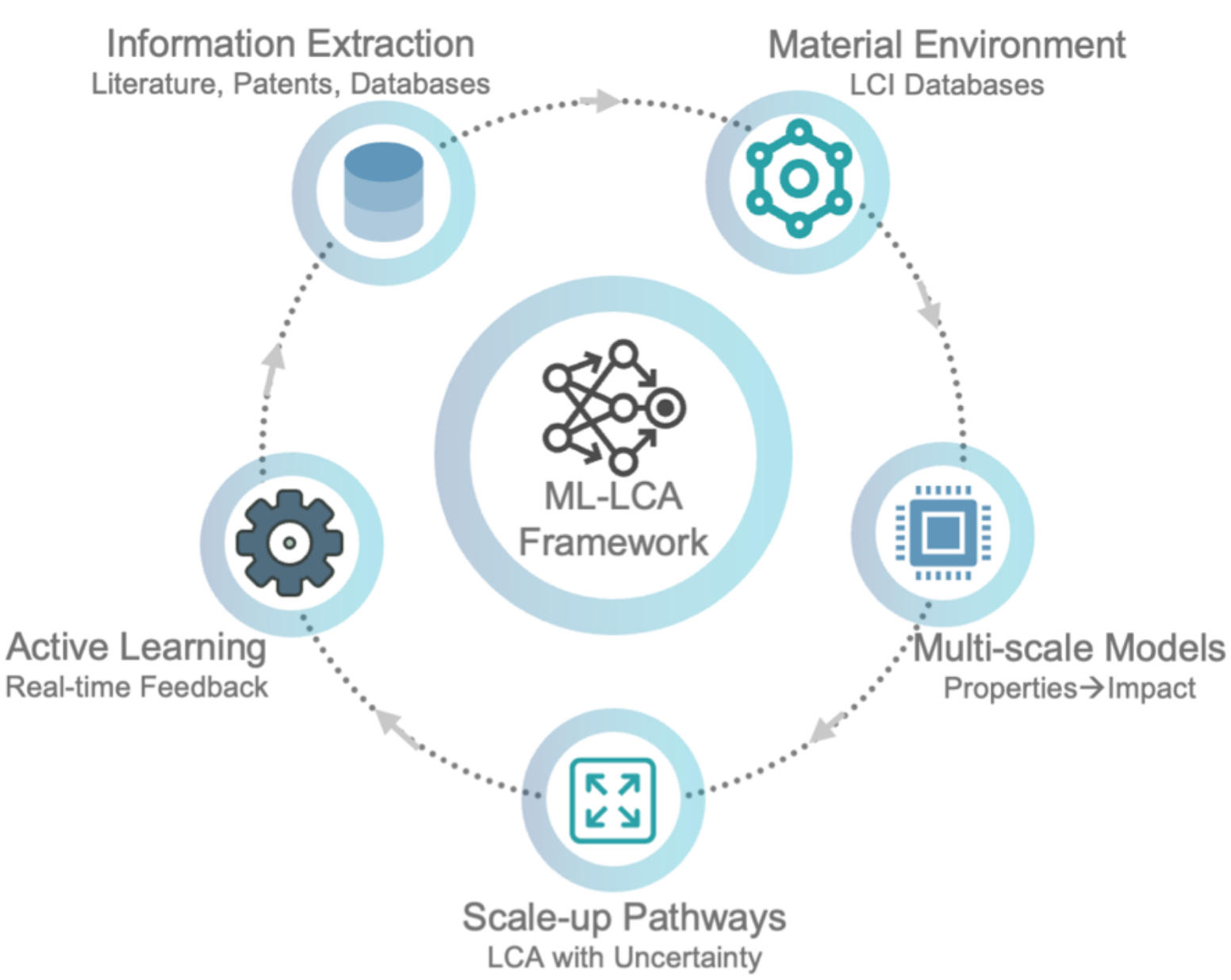

**Figure 2. The ML-LCA framework for integrated sustainable materials discovery.** Five coupled components—information extraction, materials-environment databases, multi-scale models, ensemble scale-up prediction, and uncertainty-aware optimization—enable co-optimization of functional performance and environmental impact from atomic-scale design through lifecycle deployment.

Environmental LCA quantifies a product system's performance using indicators that reflect potential impacts on ecological, social, and economic dimensions. These indicators, such as carbon footprint, resource depletion, ecosystem toxicity, etc., represent sustainability properties analogous to how strength or conductivity represent functional properties. For materials, this information can in principle be used alongside technical property data in the discovery process. Consider steel produced via the blast furnace-basic oxygen furnace route: its carbon footprint (approximately 1.8-2.0 tonnes $CO_2$ per tonne steel) complements its mechanical properties (yield strength, ductility) in design decisions.

LCA methodology continues to evolve; recent developments address nanomaterials[36] to account for hazard and microplastics[48] to account for aquatic impact with the established framework providing a sufficiently mature basis for such integration. However, for AI-discovered materials that do not yet exist, we cannot simply look up LCA data. We must develop tools capable of predicting it. The ML-LCA framework addresses these challenges through five integrated components (Figure 2) that enable co-optimization of performance and sustainability: (i) information extraction from heterogeneous sources, (ii) materials-environment databases, (iii) multi-scale predictive models, (iv) ensemble scale-up pathway prediction, and (v) uncertainty-aware optimization.

**Component 1: Information extraction and knowledge base construction**

Despite advances in research data management, only a small fraction of research data exists in usable structured form—what practitioners term “ML-ready.” Most remains embedded within scientific articles as unstructured text, representing vast untapped potential[49]. Since the early 2000s, researchers have developed dedicated extraction pipelines for chemical and scientific data[50], including rule-based approaches (ChemDataExtractor[51–53] and others[54–56]), early ML-based approaches[54–6155–6254–6153–6054–61], and hybrid approaches[62–67]. Most prior work focused on named entity recognition (NER) to construct structured databases from literature[47–52]; however, scaling across materials literature remains challenging due to heterogeneous formats, inconsistent reporting styles, and complex interrelationships between processing, structure, and properties[68].

Transformers, and more specifically LLMs, have dramatically changed this landscape. Their ability to solve tasks without explicit training makes them a powerful, scalable alternative for structured data extraction[73], enabling workflows that once required weeks of development to be prototyped in days[74]. Akin to growing reports of LLM-based extraction in chemistry[75–80] and materials science[72,81–88], substantial progress has been made in building extraction pipelines for materials and molecular data: LLM-based workflows now achieve high precision and recall on structured property extraction tasks. Such frameworks have been demonstrated for consolidating fragmented lifecycle data for specific material classes, including plastics[89], and for LCI and environmental impact data[90]. Nevertheless, a robust, general, and validated pipeline that couples materials information extraction to LCA inventory construction across the full range of inorganic, organic, and composite materials relevant to AI-driven discovery does not yet exist. The specific gap is not extraction capability per se, but the absence of an end-to-end framework that connects extracted materials data to LCA process models in a form suitable for high-throughput screening. LLMs excel at materials science data extraction because they can perform tasks such as named entity recognition and relation extraction without explicit training on domain-specific datasets, instead leveraging vast knowledge acquired during pre-training on internet-scale text[91,92]. When combined with vision capabilities for analyzing spectra and reaction schemes, multimodal capabilities, or deployed as agents that autonomously chain specialized tools, LLMs are poised to become indispensable infrastructure for building comprehensive, continuously updated materials knowledge bases at scale[73,93].

Beyond extraction pipelines, an important challenge lies in harmonizing the heterogeneous data that emerges from diverse sources. Ontology frameworks such as the Materials Ontology (MatOnto[94]), the European Materials Modeling Ontology (EMMO[95]), the PMD Core Ontology (PMDco[96]), the Materials Design Ontology (MDO[97]), and the Materials Data Science Ontology (MDS-Onto[98]) provide semantic structures for materials data, though these typically do not yet account for the multi-modality required for sustainability studies. Domain-specific extensions for life cycle assessment[99] provide semantic structures that convert inconsistent terminology and reporting conventions into machine-interoperable formats. In the future, we may increasingly see LLMs deployed to accelerate ontology construction itself (e.g., LLMs4OL[100]) by automatically identifying entity relationships from text corpora and enabling human-in-the-loop annotation workflows that reduce curation time by more than 50% compared to manual approaches[84]. By learning implicit normalization rules during fine-tuning, LLMs can map varied terminologies (e.g., "nitrogen-doped," "N-doped," "doped with N") to canonical ontology classes without extensive post-processing.

Equally important is the extraction of sustainability-relevant data that resides outside traditional peer-reviewed literature. Life cycle inventory data, embodied energy values, and supply chain information are often documented in technical reports, manufacturer specifications, safety data sheets, and regulatory filings—sources that academic extraction pipelines have largely overlooked or lack access to[90]. LLMs' ability to process long-context documents and extract structured data without domain-specific retraining makes them uniquely suited to mine this industrial and grey literature, though company-specific data silos pose challenges for public access. When integrated within a unified semantic framework, complementary data streams from scientific literature can provide fundamental structure-property relationships while industrial documentation supplies real-world processing parameters and environmental impact data, enabling truly comprehensive materials-environment databases that link molecular-level properties to global sustainability metrics.

**Component 2: Materials-environment databases**

Despite advances in materials informatics, integrating high-fidelity synthesis and environmental impact data into early-stage design remains limited. The challenge extends beyond scaling data quantity to addressing fundamental availability: ex-ante LCA predictions cannot be directly validated using industrial-scale operational data, so databases must store not only best-estimate inventory values but also scenario-based parameters, sensitivity ranges, and uncertainty descriptors that bound environmental estimates under different manufacturing assumptions. Consistency of definitions across data sources compounds this challenge; without standardized reporting of properties such as degradation rates, ecotoxicity endpoints, and recyclability metrics, ML models cannot be reliably trained across heterogeneous data. Rather than requiring comprehensive LCI databases as a precondition, the ML-LCA framework is designed to function under data-scarce conditions: surrogate LCA models trained on partial inventories guide design toward lower-impact candidates; experimental validation of those candidates generates new LCI-relevant measurements; and those measurements are used to retrain the surrogate. Each iteration expands coverage and improves accuracy, progressively building the very database the framework needs.

Recent work demonstrates that automated LCI generation is tractable at scale for organic chemicals. The CRYSTAL framework[101] uses retrosynthesis and machine-learned gate-to-gate inventories to generate consistent LCI data for more than 110,000 organic chemicals directly from molecular structure, covering feedstock and energy demands, auxiliary materials, and waste flows. A related approach, AuLCA[102], uses chemical reaction networks and first-principles energy estimation to predict LCI for fine chemicals not covered by commercial databases. These developments represent meaningful progress toward automated LCI generation for organic systems. The analogous capability for inorganic materials, such as oxides, glasses, cements, metallic alloys, does not yet exist, and extending retrosynthesis-based approaches to solid-state reaction networks and non-stoichiometric systems is a primary open challenge for Component 2.

**Component 3: Multi-scale environmental impact prediction**

Bridging the scale gap from atoms to industrial systems requires ML frameworks that translate atomic-scale structural and compositional information into estimates of lifecycle environmental impact. This demands two coupled mappings: from composition and structure to material properties, and from those properties, combined with process conditions, to lifecycle inventory data and impact indicators. Graph neural networks (GNNs) are well-suited to the first mapping, learning structure-property relationships directly from atomistic representations[103–106]. For the second, the pipeline proceeds as follows. Material properties predicted by GNNs — thermal stability, reactivity, phase equilibria — map the candidate composition to a plausible process system: high-temperature fusion for silicate glasses, reduction-based routes for metals, fermentation for bio-based polymers. Within the identified process system, relevant unit operations are enumerated with associated inventory data and emission factors drawn from LCA databases — for a metal, this spans ore preparation, reduction, refining, and forming, each route carrying different energy and emission profiles. The GNN's learned property representations then estimate scale-dependent parameters such as energy intensity, yield rates, and by-product generation, following scaling relationships established in the ex-ante LCA literature[23]. The resulting inventory is processed through LCA characterization to estimate environmental impacts, with uncertainty propagated across the full chain using ensemble methods[47]. LLMs support this pipeline by synthesizing unstructured process knowledge from literature and technical reports into structured process descriptions at the unit operation level, complementing the data infrastructure built in Components 1 and 2. Trained jointly on performance metrics and environmental indicators, such models enable co-optimization within the design loop.

Meaningful progress already exists in predicting LCA outputs from molecular structure for organic chemicals. These include neural network-based rapid LCA impact screening[107], improved data processing strategies for multi-category impact indicators[108], and ML approaches specifically for prospective LCA of emerging chemical technologies[21]. These methods are operational for cradle-to-gate assessment of organic molecules with known synthesis routes. What does not yet exist is an equivalent capability for inorganic and multi-component materials, where synthesis routes are more diverse, structure-property-process relationships are less well parameterized, and the two-step mapping, composition to process system, process system to LCI, requires physical reasoning that current models do not provide.

Multi-scale training datasets for this purpose are currently incomplete. Initial implementations may therefore focus on material classes — oxide ceramics, metallic alloys — where sufficient data exist to establish proof-of-concept, then expand as database coverage grows through the active learning loop described in Component 2. Transfer learning[3] (adapting a model pre-trained on one material class or task to a related one, reducing data requirements for the new domain) can facilitate knowledge transfer across related material families, while curriculum learning (a training strategy that introduces progressively more complex examples) can improve model robustness and generalizability. However, rigorous validation of these transfers remains essential to ensure reliability.

**Component 4: Ensemble scale-up pathways with uncertainty quantification**

Predicting how novel materials will be manufactured at an industrial scale involves substantial uncertainty. Rather than committing to a single synthesis pathway, robust assessment requires evaluating ensembles of plausible manufacturing routes and deployment contexts, each with probability distributions reflecting uncertainties in feasibility, efficiency, and cost. Machine learning models trained on existing materials databases can enable extrapolation to novel chemistries based on structural similarities and thermodynamic properties. This parallels the proxy-technology and scaling-law approaches for ex-ante LCA[23,38], adapted here for operation at the throughput and generality that ML-driven discovery demands. For a proposed material composition, similarity searches identify established materials with analogous structures or processing requirements, whose manufacturing routes serve as templates. Thermodynamic calculations constrain which routes are physically feasible (e.g., required temperatures, pressures, atmospheres). Process engineering models estimate resource consumption, emissions, and yields for each candidate route. Ensemble LCA[46,47], defined as the evaluation of environmental impacts across a distribution of plausible manufacturing routes and parameter assumptions rather than a single deterministic estimate, then evaluates the environmental profile across this distribution of possibilities, quantifying not just expected impacts but uncertainty bounds (Figure 3).

The inherent uncertainty in ex-ante LCA modeling is mitigated through this scenario-based approach: ensembles of plausible product systems, derived from AI-generated material property and synthesis information, serve as the basis for upscaling. As experimental validation or refined computational methods become available, the ensemble is refined through active learning, progressively reducing uncertainty. Maintaining transparency about these uncertainties is essential; decision-makers must understand not just predicted impacts but confidence levels to make informed choices at different development stages.

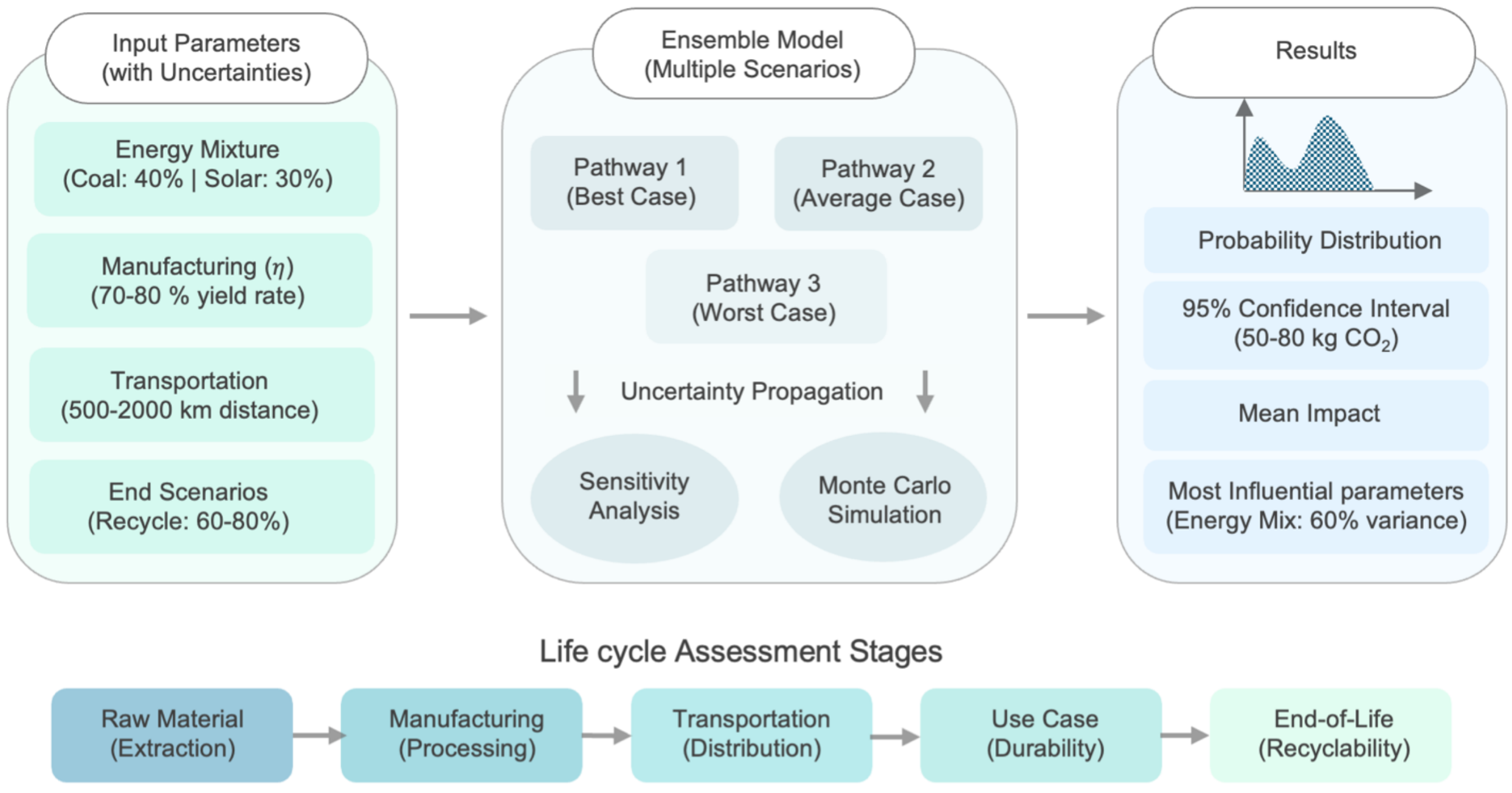

**Figure 3. Ensemble LCA pathways with uncertainty propagation mechanisms.** For a candidate material, multiple synthesis routes (solid-state reaction, sol-gel, hydrothermal, etc.) are evaluated, parameterized by probability distributions over temperature, pressure, yield, and energy consumption. Process engineering models translate these to lifecycle inventories, propagating uncertainties through to impact assessment. The result is probability distributions over environmental metrics rather than point estimates, enabling robust decision-making under scale-up uncertainty.

**Component 5: Uncertainty-aware optimization with real-time sustainability feedback**

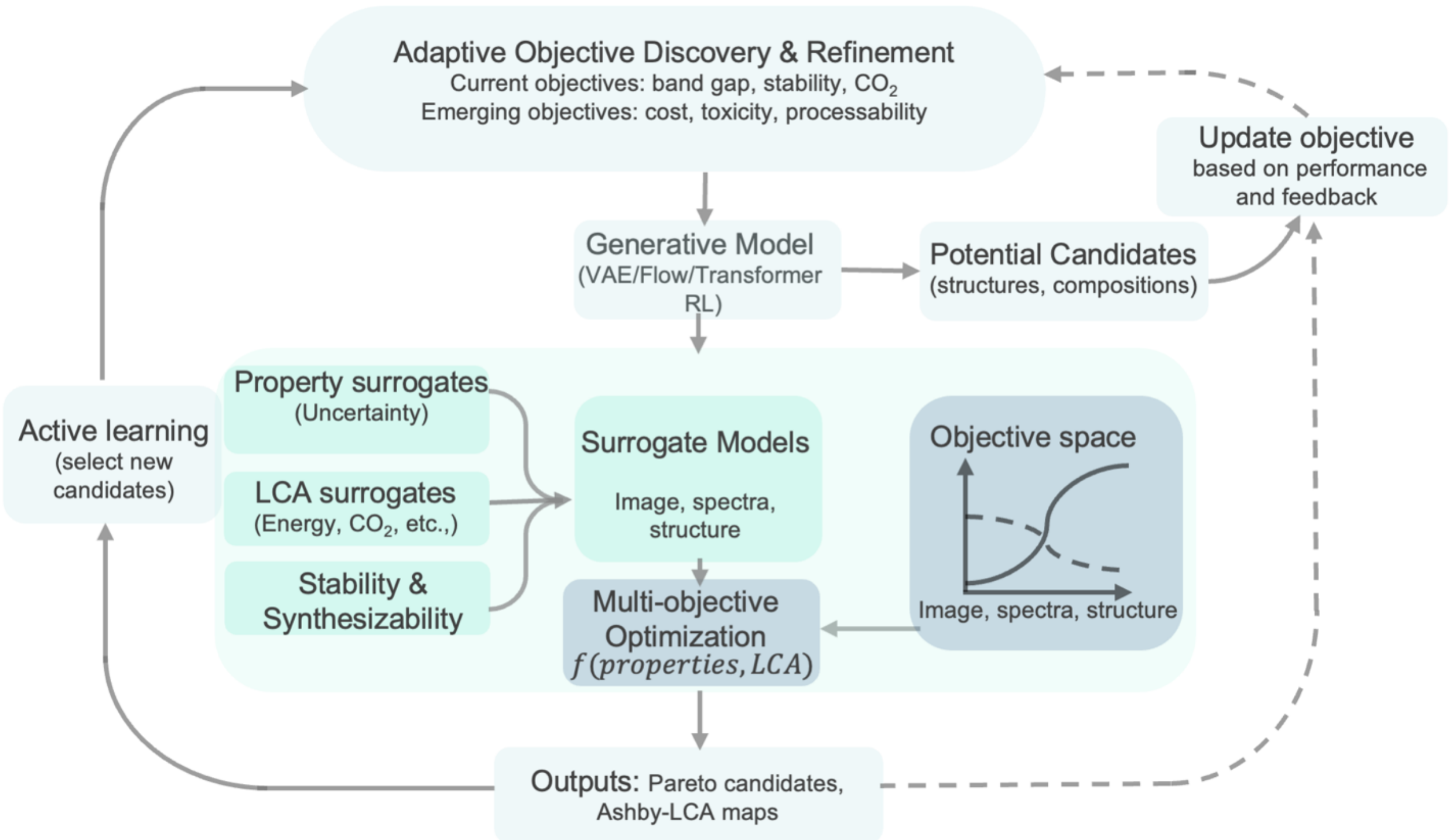

**Figure 4. AI -driven sustainable discovery with dynamically evolving design objectives.** As materials progress through the discovery pipeline from generative design to experimental validation, sustainability constraints are continuously refined. Early-stage screening uses approximate, high-uncertainty LCA estimates to filter obviously unsustainable candidates. Mid-stage assessment employs ensemble predictions across plausible synthesis routes. Late-stage validation integrates experimental data to refine lifecycle projections. Design objectives evolve from broad sustainability constraints (e.g., "carbon footprint below industry average") to specific targets informed by iterative learning.

Integrating sustainability into active learning workflows requires optimization architectures that accommodate multi-objective trade-offs, heterogeneous uncertainties, and evolving constraints (Figure 4). Sustainability-constrained optimization describes modified AI algorithms incorporating environmental constraints, including multi-objective Bayesian optimization under uncertainty, constrained neural networks with LCA objectives, and inverse design from sustainability requirements. Bayesian optimization provides a natural framework for this integration. Gaussian process surrogate models capture both expected performance and uncertainty for multiple objectives—strength, conductivity, carbon footprint, embodied energy. Acquisition functions, mathematical criteria that balance exploring uncertain regions with exploiting known high-performing ones, balance exploration (sampling uncertain regions to improve model accuracy) with exploitation (focusing on regions likely to yield good designs), explicitly accounting for uncertainty in each objective. Multi-objective formulations generate Pareto frontiers showing trade-offs between performance and sustainability, enabling stakeholders to navigate these trade-offs rather than imposing arbitrary weightings. The uncertainty bounds on these Pareto frontiers must be interpreted carefully; the conventions established for probabilistic comparative LCA[47] provide the statistical basis for determining when differences in predicted environmental impact are robust enough to inform design decisions.

A key practical question is where the training data for surrogate LCA models originate in an environment already characterized by data scarcity. The answer connects directly to the process system pipeline described in Component 3: for each candidate composition, the composition-to-process-system mapping enumerates plausible unit operations with associated process parameters and thermodynamic constraints. Process engineering models applied systematically across that parameter space generate synthetic LCI data without requiring measured industrial inventories as input. This approach, constructing surrogate training sets from process simulation rather than from empirical measurement, is established in chemical process design[109] and is directly transferable to the ex-ante LCA context. The resulting surrogate carries high epistemic uncertainty, which is intentional: at the early screening stage, the goal is not precise impact quantification but coarse ranking sufficient to filter obviously unsustainable candidates. As experimental validation data accumulate through the active learning loop of Component 2, the surrogate is retrained and its uncertainty bounds narrow progressively."

Active learning strategies determine which materials to evaluate next, prioritizing low-cost sustainability screening before committing to expensive DFT calculations or experimental synthesis. As the process progresses from early screening to late-stage validation, the strategy adapts from broad exploration to local refinement within sustainability constraints. Rather than optimizing for expected sustainability metrics, robust optimization targets worst-case scenarios within uncertainty bounds, reducing the risk of deploying materials that prove problematic under realistic conditions.

## 4. Case studies: mapping materials discovery to the ML-LCA framework

The four case studies were selected to span contrasting combinations of design-space dimensionality, data availability, sustainability driver, and ML-LCA bottleneck, rather than to be exhaustive. Polymers represent the case with a high-level of AI/ML maturity and a very broad internal design space, but also with the most heterogeneous sustainability constraints: fossil feedstock substitution, end-of-life recyclability, and ecotoxicity. Further these sustainability constraints are incommensurable and depend on which design subproblem is being addressed. Glass represents a compositionally rich inorganic system with moderate data availability where the binding bottleneck is the absence of validated structure-property surrogates for disordered materials — a model problem for Component 3. Photoresists represent the data-access-limited extreme: a technically constrained specialty chemical domain where proprietary information barriers block all five components simultaneously, illustrating the limits of the ML-LCA framework under realistic industrial conditions. Cement represents the inverse of photoresists: a data-rich domain with strong LCA foundations (Component 2) where the bottleneck is multi-scale modeling (Component 3), making it the case study where Components 4 and 5 are most tractable once that single gap is closed. Together, the four cases cover the full range of ML-LCA readiness and expose different failure modes of the framework, from data access to model architecture to integration depth. The table below summarizes the key dimensions across cases.

**Table 1. Case study selection rationale and ML-LCA readiness.** For each of the four material systems, the design space addressed, the primary sustainability driver, the data availability regime, and the binding bottleneck within the five-component ML-LCA framework.

| Material | Design Space | Primary Sustainability Driver | Data Availability | Key Bottleneck |
| --- | --- | --- | --- | --- |
| Polymer | High (monomer, backbone, additives, biobased sourcing, synthesis-route) | Fossil feedstock substitution; end-of-life recyclability | Moderate for performance; sparse for LCA | LCA data fragmented across subproblems; no closed-loop integration |
| Glass | High (multi-component composition) | Melting energy; raw material embodied carbon | Moderate for commodity glasses; absent for novel compositions | No validated surrogate for amorphous structure-property-LCA chain |
| Photoresist | Very high (molecular structure, formulation, process interactions) | PFAS persistence; solvent toxicity; process waste | Sparse; proprietary | Data access; no LCA for novel PFAS alternatives |
| Cement | Moderate (mix composition, SCM substitution) | $CO_2$ emissions from clinker production | Strong for conventional; moderate for novel binders | Multi-scale hydration models linking composition to durability and LCA |

**Polymers as a model system for ML-LCA integration**

Polymers represent one of the fastest-growing material classes globally with worldwide production reaching ~338 million tonnes in 2019, an increase of more than 640% relative to 1975[110]. They are also among the most environmentally problematic materials due to limited recyclability, poor biodegradability, and inadequate end-of-life management. Within the polymer design space, three subproblems are relevant to ML-LCA integration and differ substantially in design complexity and sustainability constraints. Monomer and backbone design, exemplified by the search for bio-based or chemically recyclable alternatives to petroleum-derived commodity plastics, involves molecular-scale generative design where the sustainability driver is primarily fossil feedstock substitution and end-of-life depolymerizability. Additive and formulation design operates at a higher level of complexity, with sustainability constraints dominated by solvent toxicity, plasticizer leaching, and processing energy. Bioplastic design, substituting petroleum-derived polymers with bio-sourced or biodegradable alternatives, involves the most heterogeneous LCA outcomes, since bio-based origin and biodegradability do not reliably reduce environmental burden without specifying feedstock, processing route, and end-of-life pathway. The discussion below treats these subproblems together where the ML/AI literature is unified and distinguishes them where sustainability constraints diverge.

AI/ML methodologies have become central to rational polymer design, with both forward (structure-to-property) and inverse (property-to-structure) paradigms reaching considerable maturity. Forward property prediction employs surrogate models based on cheminformatics descriptors, graph neural networks, or large language models, where polymer repeat units or oligomeric fragments encoded as SMILES strings are mapped to target properties[111]. Multimodal ML frameworks integrating multiple representation schemes—combining GNNs, LLMs, and chemical descriptors—demonstrate improved predictive accuracy and transferability[112,113]. Inverse polymer design has advanced through generative strategies including genetic algorithms[114], variational autoencoders[115], recurrent neural networks[116], and diffusion models[117], often coupled with pretrained property predictors. Open-access polymer databases[118–120] spanning homopolymers, copolymers, and diverse physicochemical properties have been critical in enabling the development and validation of these approaches.

Success stories demonstrate the potential: polymers designed for high-temperature dielectrics for energy storage[121], star block copolymers with ultralow interfacial tensions[122], and sustainable packaging materials[123]. Although limited, some examples target sustainability explicitly rather than performance alone. Genetic algorithms and multitask neural networks have enabled the design of bio-synthesized and biodegradable polyhydroxyalkanoates (PHAs), promising alternatives to petroleum-based commodity plastics[124,125]. A complementary approach[126] targets chemically recyclable polymers via ring-opening polymerization (ROP), using GA-based inverse design coupled with surrogate ML models to screen millions of candidates for depolymerizable chemistries.

A key limitation of current AI/ML-driven polymer design is the implicit assumption that bio-based, biodegradable, or chemically recyclable polymers are inherently sustainable. This assumption often fails under full lifecycle scrutiny. While bio-based polymers reduce reliance on fossil resources, their overall environmental footprint depends strongly on feedstock sourcing, processing energy, end-of-life pathways, and integration into existing waste management infrastructure. Rigorous LCA is therefore essential. For example, LCA studies indicate that bio-based polylactic acid (PLA) can have higher environmental impacts than petroleum-derived polyethylene terephthalate (PET)[127,128]. Consequently, biodegradability and bio-based origin cannot serve as sustainability proxies.

Owing to high costs and extensive data requirements across production-use-reuse-recycling value chains, comprehensive LCA studies exist for only a limited set of polymers: PLA, bio-PET, PBS, bio-PE, SBS, cellulose acetate[127,129]. The value of these studies extends beyond verification; they identify process stages contributing substantially to greenhouse gas emissions, enabling targeted optimization. For instance, LCA of PLA reveals that biomass conversion to lactic acid followed by polymerization is highly energy-intensive, contributing substantially to $CO_2$ emissions[130]. Process refinement targeting these hotspots offers significant benefits. A key outcome across diverse polymer LCA studies is that bioplastics derived from renewable or waste-based resources, coupled with appropriate end-of-life routes such as anaerobic digestion, represent promising directions—but require clear regulatory frameworks and economic incentives for large-scale adoption.

**Integrating AI/ML with LCA for polymers**: Polymers offer an ideal testbed for ML-LCA integration given the relatively mature state of both AI-driven design and lifecycle assessment. We outline strategies for integration:

(a) **Direct LCA prediction**: One approach learns LCA outputs—GHG emissions or sustainability indicators—directly from existing polymer LCA databases. Admittedly, accuracy will be limited by data scarcity and uncertainty in processing conditions for novel systems. However, the intent is not replacing full LCA but enabling low-fidelity screening, followed by rigorous assessment for promising candidates.

(b) **Simplified sustainability indicators**: Since direct LCA learning is challenging, the problem can be reformulated to predict simplified metrics. Recent work[131] used ML to identify greener solvent alternatives by learning G-scores (a sustainability metric from the GlaxoSmithKline Solvent Sustainability Guide). Gaussian process regression models trained via cheminformatics estimated greenness of over 10,000 solvent candidates, identifying novel replacements for 29 hazardous solvents including benzene and diethyl ether.

(c) **LCA-tool-generated training data**: Transformer-based models[132] predict product carbon footprints (PCFs) directly from molecular SMILES representations. Notably, ground-truth PCF data were generated using SimaPro LCA[133] with the IPCC[134] 2013 impact assessment method. This exemplifies using LCA tools to generate proxy sustainability labels for ML training, enabling rapid initial screening of millions of candidates. The approach extends to multiple impact indicators—acidification, human health, ecosystem quality, particulate matter formation—capturing environmental sustainability across multiple dimensions[21,107,108].

(d) **Prospective forecasting**: ML offers opportunities for forecasting future sustainability trends[21]. Key factors such as resource depletion and water scarcity are strongly time- and location-dependent; regions unaffected today may face significant constraints in the future. ML-based forecasting of environmental conditions, integrated with LCA frameworks, enables understanding how environmental impacts may evolve. ML methods can also aid global sensitivity analysis in LCA[135].

**ML-LCA framework gap analysis.** Component 1 is moderately developed: polymer structure and property data are reasonably abundant, but sustainability-relevant quantities (biodegradation rates under different environmental conditions, ecotoxicity endpoints for monomers and additives, end-of-life fate fractions across recycling, incineration, and landfill pathways) remain scattered across inconsistent reporting formats and are largely absent from ML-ready databases. Component 2 has growing but thin LCA coverage: comprehensive inventory data exist for a handful of well-studied biopolymers (PLA, bio-PET, PBS, cellulose acetate) but are essentially absent for the novel chemistries that generative design is now producing, such as sequence-defined recyclable polymers and biosynthetic PHAs with non-standard monomer compositions. Component 3 is partially operational but unevenly so: forward property prediction from repeat-unit structure is mature for different thermal, electronic, mechanical or optical properties, but linking molecular structure to processing energy, synthesis route emissions, and end-of-life behavior — the quantities LCA requires — remains nascent and largely undeveloped. The LCA-tool-generated training data approach (predicting product carbon footprints from SMILES) is a promising step but covers only cradle-to-gate impacts and a narrow set of impact categories. Component 4 is underdeveloped: virtually all published polymer design work assumes a single, fixed synthesis route rather than a distribution of plausible routes with associated uncertainties; the ensemble pathway logic of Component 4 has not been demonstrated for any polymer system. Component 5 exists only for performance objectives; no published example closes the loop by using LCA-derived constraints to guide polymer inverse design, which represents the core of what ML-LCA is intended to achieve.

**Path forward.** Polymers offer a tractable proving ground for ML-LCA integration. The combination of mature AI models, reasonable data availability, and societal urgency around plastic waste creates an opportunity for rapid progress. Key research directions include expanding LCA databases with standardized protocols, developing multi-fidelity screening models, incorporating end-of-life uncertainty through ensemble predictions, and demonstrating closed-loop optimization where sustainability constraints guide discovery alongside performance targets.

### Glass: Data scarcity limits AI-LCA integration

Glass is used extensively in architecture, electronics, packaging, and renewable energy—from photovoltaic panel covers to reinforcement fibers in wind turbine blades—playing a vital role in modern infrastructure. Despite its ubiquity, the glass industry remains highly energy-intensive, contributing substantially to global carbon emissions. While soda-lime silicate glass dominates the market, the discovery and optimization of novel glass compositions for tailored properties—optical clarity, thermal resistance, chemical durability—still rely heavily on empirical formulation.

One key limitation in developing surrogate AI models for glassy systems is the lack of high-quality, standardized datasets for amorphous materials[136]. Most available data are sparse, system-specific, and lack the compositional and structural diversity needed to train robust predictive models. Although LCA data exist for various glass types and processing pathways, these environmental metrics are rarely integrated into forward or inverse materials design workflows[137–140]. Moreover, existing AI-driven discovery methods struggle to accurately capture the disordered nature of amorphous structures, and there is currently no standardized surrogate modeling framework that connects glass composition, thermal history, and processing parameters to macroscopic performance properties[136].

**ML-LCA framework gap analysis.** Component 1 is partially operational: extraction pipelines for glass composition-property data have been demonstrated[64], but sustainability-relevant quantities (melting energy as a function of batch composition and cullet fraction, furnace type and throughput, emissions from carbonate decomposition) remain largely unstructured in the literature and absent from existing databases. Component 2 has reasonable LCA coverage for commodity glass types (soda-lime-silicate flat glass, container glass), where ecoinvent entries provide documented inventory data. For the novel multicomponent compositions targeted by AI-driven discovery—such as borosilicates, chalcogenides, fluoride glasses, and high-entropy oxide systems—no lifecycle inventory data exist, and proxy technology selection is poorly constrained because processing routes differ substantially from commodity glass in energy intensity and emission profile. Component 3 is the binding constraint for the entire framework: no validated surrogate model connects glass composition and thermal history to the macroscopic properties — viscosity curve, optical constants, chemical durability, mechanical strength — that govern both performance and processing energy. Without this, the composition-to-process mapping required by Component 3 cannot be executed and the scale-up parameterization of Component 4 has no property inputs to work from. Component 4 is therefore non-operational for novel compositions; the proxy technology approach is feasible only within well-characterized composition families such as soda-lime-silicate, where the process system is known. Component 5 is similarly blocked. The gap between current capability and the ML-LCA ideal is, for glass, almost entirely a Component 3 gap — one of model architecture and training data for amorphous structure-property relationships — and closing it is a prerequisite before any other component can meaningfully advance.

**Path forward.** Establishing large, shared datasets of glass compositions, processing conditions, measured properties, and LCA outcomes would enable training integrated surrogate models. Recent efforts to develop glass-specific descriptors and neural network potentials for amorphous systems offer promising starting points. Coupling these with process-based LCA models could enable screening novel glass compositions for both performance and environmental impact before expensive synthesis trials.

**Photoresists: Proprietary data hinders sustainable PFAS alternatives**

While per- and polyfluoroalkyl substances (PFAS) offer exceptional chemical stability, thermal resistance, and hydrophobicity, their environmental persistence prompts urgent efforts to identify replacements, increasingly using AI/ML approaches[141,142]. Semiconductor manufacturing presents one of the most technically demanding cases for PFAS replacement. In extreme ultraviolet (EUV) and deep ultraviolet (DUV) lithography, chemically amplified resists rely on fluorinated photoacid generators (PAGs) and polymer backbones (e.g., hexafluoroisopropyl groups) that control acid diffusion, enhance EUV absorption, and enable sub-5 nm patterning resolution, performance attributes with virtually no fluorine-free equivalents demonstrated at commercial scale[143]. The challenge intensifies because photoresist formulations are highly process-specific; minor formulation changes require extensive reformulation, testing, and yield requalification across numerous device-specific recipes[144].

Recent work, ParetoGen[145], uses generative ML algorithms to identify chemicals on the Pareto optimal frontier of multiobjective optimization, simultaneously maximizing functionality and minimizing chemical hazard for PFAS green substitutes. However, hazard in ParetoGen is defined as chemical toxicity and persistence, not lifecycle environmental impact. A candidate identified as low-hazard by this metric may still carry high embodied energy in synthesis, generate fluorinated precursor waste streams, or lack viable end-of-life pathways. The ML-LCA gap for photoresists is therefore not in molecular design capability, which is advancing, but in the absence of lifecycle-aware optimization: no published approach integrates LCA impact estimates alongside lithographic performance and chemical hazard as joint objectives in the design loop.

**ML-LCA framework gap analysis.** Component 1 is severely hampered: photoresist formulations are proprietary and performance metrics reside in manufacturer qualification reports rather than peer-reviewed literature. Component 2 lacks lifecycle data for novel PFAS alternatives; while recent work has substantially advanced characterization of PFAS waste streams in semiconductor manufacturing[146] and transformation pathways of fluorinated lithographic materials[147], full cradle-to-gate lifecycle inventories for alternative PFAS-free resist formulations, covering synthesis energy, precursor burdens, and end-of-life, remain absent. Component 3 faces the challenge of linking molecular properties (acid diffusion, optical absorption) to lithographic performance (resolution, line-edge roughness) and then to environmental metrics, a chain requiring extensive proprietary process knowledge that is not publicly available. The work of ParetoGen demonstrates that the molecular design to hazard step is tractable on public data; it is the hazard-to-LCA extension and the performance-to-process step that remain blocked. Components 4–5 are non-operational without Component 1–3 foundations.

**Path forward.** ParetoGen's Pareto frontier framework is a natural starting point for ML-LCA integration: the same multiobjective optimization infrastructure that balances functionality against chemical hazard could, in principle, incorporate LCA impact indicators as additional objectives once inventory data become available. This requires two developments: pre-competitive data sharing between manufacturers and researchers to generate the LCI data that Component 2 currently lacks, and extension of Component 3 modeling to map molecular structure to process-level LCA parameters alongside lithographic performance. Industry consortia that protect competitive formulation details while enabling aggregate sustainability assessment represent the most tractable near-term pathway.

**Cement: Rich LCA data, limited integration with materials design**

Cement is among the most widely used construction materials globally and also among the largest contributors to $CO_2$ emissions, accounting for approximately 7-8% of global carbon emissions[148]. Despite decades of industrial development, the cement sector continues to depend heavily on empirical trial-and-error methods for process and mix optimization. Strategies such as reducing clinker content, optimizing mix-design parameters, adjusting kiln process parameters, or exploring alternative pozzolanic activation pathways (e.g., mechanochemical activation) can substantially decrease $CO_2$ emissions. However, robust predictive models linking mix design parameters to both durability and carbon footprint remain lacking, hindering systematic trade-off assessments.

Paradoxically, cement benefits from relatively rich LCA data compared to emerging materials—decades of environmental analysis have documented the lifecycle impacts of Portland cement production, clinker substitution strategies, and alternative binders. Yet these data remain disconnected from materials-level optimization. The hydration chemistry, microstructure development, and long-term durability of cement are governed by complex, multi-scale phenomena—dissolution kinetics of clinker phases, nucleation and growth of hydration products, pore structure evolution—that current AI models struggle to capture comprehensively.

**ML-LCA framework gap analysis.** Component 1 is reasonably developed; extensive literature exists on cement chemistry, though extracting structured data requires substantial effort. Component 2 is strong for conventional cement but weaker for novel formulations (geopolymers, alkali-activated materials, ultra-low clinker cements). Component 3 represents the primary bottleneck—predicting mechanical properties and durability from composition and processing requires coupling thermodynamic models (phase assemblages), kinetic models (hydration rates), and microstructural simulations (pore structure), which current AI surrogates handle incompletely. Components 4-5 could build on the strong LCA foundation if multi-scale prediction gaps were addressed.

**Path forward.** AI-driven frameworks connecting molecular features (e.g., calcium silicate reactivity), hydration dynamics, microstructural evolution, and LCA metrics remain lacking but represent a tractable research direction given the extensive existing knowledge base. Integrating thermodynamic databases (e.g., Cemdata[149]) with machine-learned models of hydration kinetics and microstructure-property relationships could enable closed-loop, data-driven sustainable cement design—optimizing simultaneously for performance (strength, durability) and environmental impact (embodied carbon, resource efficiency).

## 5. Outlook

The path forward is challenging, demanding bridging of fundamental scale gaps between atomistic simulations and industrial processes, development of ex-ante LCA methodologies for materials that do not yet exist, management of compounding uncertainties across prediction stages, and creation of data infrastructure integrating materials science with environmental assessment. Yet the potential impact is substantial: enabling AI to discover materials that are genuinely sustainable by design, rather than discovering high-performance materials and hoping they prove sustainable in retrospect. Realizing this vision requires coordinated advances in data infrastructure, multi-scale modeling, ex-ante LCA methodology, and uncertainty-aware optimization, with industrial adoption timescales ranging from 5–10 years for regulatory-driven sectors to 10–20 years for others. The ML-LCA framework developed here addresses environmental impact prediction and optimization; social and economic sustainability dimensions, while important, require complementary assessment methodologies such as social LCA and techno-economic analysis towards which the approach could be extended. Several challenges exist beyond technical development to encompass data access, workforce training, validation of prospective predictions, and regulatory alignment (see Supplementary Section 5 for a detailed discussion). Ensuring these materials are sustainable by design is among the most consequential scientific and engineering challenges of the 21st century.

## Author contributions

All the authors conceived the idea jointly through discussions, wrote the first draft of the paper, reviewed and approved the final manuscript.

## Competing interests

The authors declare no competing interests.

## Data availability

No new data were generated for this perspective article.

## Acknowledgments

NMAK acknowledges the funding support from Anusandhan National Research Foundation, and Google Research Scholar Award. SM acknowledges the funding support from the Prime Minister’s Research Fellowship. RB acknowledges funding support from Anusandhan National Research Foundation. LW acknowledges funding support from the Carl Zeiss Foundation. RM acknowledges funding support from the Wallenberg AI, Autonomous Systems, and Software Program (WASP), supported by the Knut and Alice Wallenberg Foundation.

# SUPPLEMENTARY INFORMATION

### Supplementary Section 1: Materials discovery and sustainability landscape

Materials discovery and sustainability assessment operate in parallel, rarely intersecting until late-stage deployment. Materials design optimizes atomic and microstructural properties through combinatorial experimentation, quantum calculations, and ML-predicted structure–property relationships, while LCA quantifies environmental burdens at macroscopic scales relevant to industrial processes and supply chains. This scale mismatch extends to data: materials databases contain millions of computed structures; LCA databases document thousands of characterized industrial processes. Environmental hotspots identified through retrospective LCA therefore arrive too late to inform early-stage design decisions.

The gap is real but not without partial precedent. Supply chain risk metrics such as the Herfindahl-Hirschman Index (HHI)[1,2], which quantifies the concentration of raw material production among supplier countries (near zero for diversified supply; 10,000 for single-country monopoly), have been used to flag geopolitical vulnerabilities in materials selection. However, these metrics operate outside standard LCA frameworks and are not integrated into generative discovery pipelines. More broadly, materials scientists lack feedback on how atomic-scale choices propagate to lifecycle burdens, while LCA practitioners lack frameworks for materials that do not yet exist.

### Toward circular-by-design integration

Closing this gap requires embedding sustainability throughout discovery rather than appending it retrospectively. Four integration points define this paradigm.

a. **Standardized data exchange**: Materials discovery and LCA should communicate through common ontologies enabling seamless translation between atomistic descriptors (composition, structure, bonding) and lifecycle parameters (energy intensity, precursor requirements, waste generation), facilitating automated information flow from generative models to LCA estimation.

b. **Real-time feedback**: Rather than awaiting industrial-scale data, lifecycle implications should be estimated during early design. As generative models propose candidates and surrogate models predict properties, parallel LCA estimation provides real-time sustainability feedback, enabling rapid filtering before expensive synthesis validation.

c. **Predictive modeling across scales**: AI surrogates must bridge scale gaps, predicting how atomic properties determine manufacturing processes, how processes scale to industrial production, and how industrial systems integrate into global supply chains. Though uncertain, these predictions enable prospective assessment guiding rather than merely evaluating design decisions.

d. **Traceability and trust**: Integrated frameworks must maintain transparency about data sources, model assumptions, and uncertainty levels. Stakeholders need to understand the provenance and reliability of sustainability predictions to make informed decisions—requiring interpretable frameworks explaining how design choices connect to environmental outcomes.

## Supplementary Section 2: LCA methodology: a primer for materials scientists

Life cycle assessment emerged in the late 1960s from practical industry needs, beginning with the Coca-Cola Company's analysis of beverage packaging to compare environmental impacts of different container materials[3]. Early studies focused primarily on energy consumption and solid waste, gradually expanding to encompass broader environmental impacts. The methodology gained prominence in the 1990s through two critical developments: establishment of the *International Journal of Life Cycle Assessment* as a dedicated scientific forum, and publication of the ISO 14040 series standards[4–6], which provided internationally recognized guidelines for conducting and reporting LCA studies. These standards formalized the four-stage framework that remains the methodological foundation today.

These standards formalize a four-stage framework: (i) goal and scope definition, (ii) life cycle inventory (LCI) analysis, (iii) life cycle impact assessment (LCIA), and (iv) interpretation. For materials, the functional unit — the quantified performance function that the product system delivers, for example "1 tonne of structural concrete with 28-day compressive strength ≥ 30 MPa" — anchors all subsequent calculations and governs comparability across studies. System boundaries define which lifecycle stages are included: cradle-to-gate (extraction through manufacturing), cradle-to-grave (extending through use and end-of-life), or cradle-to-cradle (additionally accounting for recycling and reuse).

Inventory analysis compiles all material and energy flows crossing the system boundary. The foreground system covers direct processes under study; the background system covers indirect supporting processes (electricity generation, transport) drawing on databases such as ecoinvent.

Elementary flows — emissions to air, water, and soil; resource extractions; land use — become inputs for impact assessment. Impact assessment translates these inventory results into indicators reflecting potential damage across categories including climate change (global warming potential, in kg $CO_2$-eq), resource depletion, ecosystem quality, and human health toxicity, using characterization methods such as ReCiPe, TRACI, or IMPACT World+[7]. Interpretation synthesizes results through contribution analysis (identifying dominant lifecycle stages), sensitivity analysis, and uncertainty quantification.

For emerging materials, the central challenge is that inventory construction requires substantial estimation: unit process data must be projected from laboratory-scale demonstrations or analogous technologies, and allocation — attributing environmental burdens when processes produce multiple products — becomes particularly challenging when synthesis pathways are uncertain[8,9]. This is the domain of ex-ante LCA, discussed in Section 3 below.

Environmental LCA addresses the ecological and resource dimensions of sustainability. It is complemented by techno-economic analysis (TEA), which quantifies costs using largely the same process models, and by social LCA, which extends lifecycle thinking to worker health, fair wages, and community impacts. These methods are briefly noted in the main text Outlook; readers are directed to the dedicated literature for methodology[8].

**Supplementary Section 3: Ex-ante and prospective LCA: methodological foundations**

**The standard–prospective distinction.** Standard LCA is retrospective: it characterizes the environmental impacts of technologies that exist at industrial scale, drawing on measured inventory data. Ex-ante LCA — also termed prospective LCA in recent terminological usage[10,11] — shifts this to technologies that have not yet reached commercial maturity, using estimated or projected inventory data derived from laboratory or pilot demonstrations. Cucurachi et al. (2018)[12] define ex-ante LCA studies as those that either (i) scale up an emerging technology using likely scenarios of future performance at full operational scale, or (ii) compare an emerged technology at scale with the evolved incumbent technology. Buyle et al. (2019)[13] propose a generic framework spanning the full technology life cycle from early design through market penetration, and categorize practical methods for incorporating future-oriented features. Arvidsson et al. (2024)[11] provide the most recent terminological synthesis, recommending harmonization around "prospective LCA" and emphasizing technology readiness level (TRL) as the key variable that determines which upscaling method is appropriate at a given development stage. The field has also developed regulatory guidance: the Safe and Sustainable by Design (SSbD) framework adopted by the European Chemicals Agency explicitly requires ex-ante assessment at early development stages.

**Upscaling methods.** Tsoy et al. (2020)[14] review upscaling methods in ex-ante LCA, identifying three main approaches:

*Process simulation.* Engineering models estimate material and energy flows for a hypothetical production system. This approach is most applicable when the synthesis route type is known, and process engineering relationships (mass balances, energy requirements, thermodynamic constraints) can be parameterized from laboratory data or first principles. Examples include prospective LCA of electrochemical processes and $CO_2$ reduction technologies, where reactor-scale models provide the primary inventory inputs.

*Molecular structure models (MSMs).* Inventory parameters are derived from molecular-level properties using established scaling relationships between molecular composition and thermodynamic or kinetic quantities. This approach is most applicable to organic chemicals with well-characterized group-contribution methods. Recent ML extensions have automated this step at scale: the CRYSTAL framework[15] uses retrosynthesis and machine-learned gate-to-gate inventories to generate LCI data for more than 110,000 organic chemicals directly from molecular structure, and AuLCA[16] uses chemical reaction networks and first-principles energy estimation to predict LCI for fine chemicals not covered by commercial databases. These represent the current frontier for organic systems; the analogous capability for inorganic and multi-component materials does not yet exist.

*Proxy technologies.* Inventory data from an established analogous technology serve as a starting estimate, adjusted for compositional or process differences. Proxy selection requires domain judgment: a novel fluoride glass composition might use soda-lime-silicate flat glass production as a first approximation, then correct for the higher melting temperature and batch complexity. The quality of this approximation degrades rapidly as compositional novelty increases. Erakca et al. (2024)[17] provide a systematic review of how scale-up method selection should be driven by TRL: process simulation is most reliable at TRL 4–6 (laboratory to pilot scale), while proxy technologies dominate at TRL 1–3 (conceptual and early laboratory stages).

**The functional unit problem for AI-discovered materials**

Ex-ante LCA for emerging technologies typically assumes a stable functional unit — 1 kWh of electricity generated, 1 $m^2$ of photovoltaic panel — that holds across TRL levels. For AI-discovered materials, the functional unit may itself be uncertain: the same composition may serve different applications with different durability requirements, use-phase intensities, and end-of-life pathways, each implying a different lifecycle profile. A high-entropy oxide glass discovered by generative screening could serve as a packaging material, an optical component, or a biomedical substrate — with radically different functional units and dominant impact categories in each case. This ambiguity is not resolvable at the computational screening stage and must be acknowledged as a source of scenario uncertainty in early-stage ML-LCA estimates. The ML-LCA framework addresses this by treating functional unit choice as part of the design specification rather than a fixed constraint, allowing the optimization to reflect the application-dependent sustainability profile of candidate materials.

**Specific challenges posed by AI-discovered materials**
Conventional ex-ante LCA assumes that the technology type, however early-stage, is defined. AI-driven generative models remove even this constraint, proposing compositions with no industrial precedent. Four challenges go beyond those addressed in standard ex-ante practice.

*Compositional novelty.* Generative models can propose element combinations or structural motifs outside the chemical space of existing industrial processes. When no analogous proxy technology exists, process modeling must rely entirely on first principles, compounding uncertainty well beyond the ranges characterized in the ex-ante LCA literature.

*Synthesis route multiplicity.* A given composition may be synthesizable through numerous pathways — solid-state reaction, sol-gel, hydrothermal, chemical vapor deposition — each with vastly different energy requirements and emission profiles. Predicting which route will prove industrially viable at the generative design stage is fundamentally uncertain, requiring ensemble treatment rather than commitment to a single pathway.

*Property–process coupling.* A material's lifecycle impact depends not only on composition but on processing parameters that determine microstructure and properties. The same composition processed differently may yield different performance and different environmental burdens, meaning that LCA inputs and property predictions cannot be decoupled.

*Throughput requirements.* Generative discovery evaluates thousands or millions of candidates in a single campaign. Full LCA for each candidate is computationally and methodologically infeasible. This demands surrogate LCA models capable of providing sufficient accuracy for coarse ranking at early screening stages, with full prospective assessment reserved for shortlisted candidates — a tiered fidelity strategy that has no direct precedent in the ex-ante LCA literature.

These challenges motivate the ML-LCA framework proposed in this Perspective: Components 3 and 4 specifically target compositional novelty (via composition-to-process-system mapping) and synthesis route multiplicity (via ensemble pathway evaluation), while Component 5 addresses throughput requirements through surrogate LCA models built from process simulation rather than measured industrial inventories.

**Supplementary Section 4: AI-driven materials discovery: landscape and sustainability gap**

**From trial-and-error to high-throughput computation**. Materials discovery has traditionally been driven by trial-and-error approaches guided by domain expertise, with researchers proposing candidate materials and validating them through laboratory synthesis and characterization. Computational methods, principally density functional theory (DFT)[18], significantly accelerated this process by enabling prediction of structural, electronic, and thermodynamic properties prior to synthesis[19]. High-throughput DFT screening, now operating at scales of millions of calculations through databases such as the Materials Project[20], the AFLOW repository, and the Open Quantum

Materials Database, has identified candidate materials for batteries, catalysts, thermoelectrics, and piezoelectrics at a pace that experimentalists cannot match[21]. Yet DFT screening remains limited to known structure types, is computationally demanding for large unit cells, and provides no direct information on synthesizability or lifecycle impact.

**The generative model revolution.** The recent introduction of generative AI models has qualitatively changed the materials discovery paradigm by enabling de novo design of crystal structures and molecular architectures rather than screening known ones. Variational autoencoders (VAEs)[22], diffusion models[23,24], flow-based models[25], and equivariant graph neural networks now operate across the full space of inorganic crystal structures, proposing stable candidates conditioned on target properties such as band gap, bulk modulus, or ionic conductivity. MatterGen[24] demonstrated diffusion-based generation of inorganic crystals with substantially improved stability and diversity relative to earlier models, and introduced property-conditioned generation as a practical design tool. GNoME[26] extended screening to over 2 million previously undocumented stable inorganic crystals by combining generative exploration with large-scale DFT stability filtering. Autonomous experimental platforms, including the A-Lab[27], close the loop from computational prediction to physical synthesis, executing and evaluating synthesis routes without human intervention between steps.

For molecular systems, diffusion models (e.g., DiffSBDD[28], EDM[29]) and flow-based models have achieved state-of-the-art performance on molecular generation benchmarks, while multimodal foundation models integrate structural, spectroscopic, and property information for improved cross-domain transferability. Machine-learned interatomic potentials (MLIPs) — including MACE[30], CHGNet[31], and SevenNet[32] — extend this landscape to dynamical properties and phase stability, enabling ab initio-quality molecular dynamics at a fraction of DFT cost.

**Coverage gaps and the sustainability blind spot.** Despite this progress, the generative model landscape has three important limitations documented in recent benchmarking. First, generative coverage remains concentrated on crystalline inorganic materials: models for amorphous systems (glasses, polymers) are substantially less mature. Second, synthesizability prediction lags structural generation: LeMat-GenBench[33] evaluates state-of-the-art crystal generative models and notes that synthesis feasibility "remains difficult" and lifecycle integration is acknowledged as a future challenge rather than a current capability. Third — and central to this Perspective — sustainability assessment is entirely absent from the design objective, training data, and candidate filtering of every major generative model currently in use. This is not a minor omission. It means that the materials entering experimental synthesis pipelines from AI-driven discovery campaigns carry no lifecycle information, and that the investment in autonomous synthesis infrastructure is being directed toward candidates selected purely for structural and functional merit.

The disconnect is not inherent to the generative modeling approach: it reflects the current absence of ML-ready environmental data and automated prospective LCA tools compatible with the throughput and generality that these models require. Closing this gap is the specific contribution of the ML-LCA framework. Each of the five components in the framework (Section 3 of the main text) addresses one of the bottlenecks identified in this section: Component 1 targets the absence of ML-ready LCA data; Component 2 addresses database coverage and standardization; Component 3 provides the composition-to-process mapping that generative models currently lack; Component 4 introduces ensemble treatment of synthesis route uncertainty; and Component 5 integrates these capabilities into the optimization loop.

**Regulatory and competitive pressure as adoption drivers.** The transition to sustainability-integrated AI discovery is accelerating under external pressure from two directions. The European Union's Digital Product Passport (DPP) initiative mandates lifecycle information for products in regulated sectors, beginning with batteries, electronics, and textiles, requiring manufacturers to document material composition, environmental footprint, recyclability, and supply chain transparency. For materials discovered by AI, this creates a compliance requirement that extends upstream into the design stage. Separately, the Safe and Sustainable by Design (SSbD) framework requires that environmental assessment be integrated at early development stages rather than deferred to scale-up, directly incentivizing the kind of closed-loop ML-LCA integration proposed here. Early adoption therefore offers both a competitive advantage and a reduced risk of costly redesigns when regulatory requirements materialize.

## Supplementary Section 5: Outlook: challenges beyond technical development

Realizing the ML-LCA vision requires progress on several fronts that extend beyond the technical framework described in the main text.

*Data access and pre-competitive sharing.* The most severe data bottleneck for photoresists and specialty chemicals is not a modeling problem but a governance one: lifecycle inventory data for novel PFAS alternatives and advanced semiconductor materials exist within companies but are not publicly available. Pre-competitive consortia that pool sustainability data while protecting commercial formulation details offer the most tractable near-term pathway for these material classes.

*Workforce and community capacity.* Implementing ML-LCA frameworks requires practitioners fluent across materials informatics, process engineering, and LCA methodology — a combination that currently exists in very few groups. Curriculum development bridging these communities and targeted training programs at the intersection of AI for science and industrial ecology represent a structural requirement for adoption at scale.

*Validation of prospective predictions.* The framework's predictions are prospective by construction and cannot be validated at the time they are made. Establishing validation protocols — comparing early-stage ML-LCA estimates against ex-post LCA data as materials mature to commercial

production — is essential for calibrating confidence and identifying systematic biases. Longitudinal tracking of materials from AI-generated candidate to industrial production would provide the empirical basis for this calibration.

*Regulatory alignment.* LCA methodology and impact characterization continue to evolve, with ongoing development of frameworks for nanomaterials, microplastics, and novel synthetic chemicals. ML-LCA surrogates trained on current LCA databases will require periodic retraining as impact characterization methods update. Alignment between the ML-LCA framework and evolving regulatory guidance, including the SSbD framework and the Digital Product Passport, will require active coordination between the research community and standards bodies.

## Supplementary Section 6: Glossary of key terms

Terms are grouped by community of origin. Definitions are given in the context in which each term is used in this manuscript.

### LCA and sustainability

*Life Cycle Assessment (LCA):* A methodology for quantifying the environmental impacts of a product or process across its full life cycle, from raw material extraction through manufacturing, use, and end-of-life disposal, following the four-stage framework of ISO 14040/14044: goal and scope definition, life cycle inventory (LCI), life cycle impact assessment (LCIA), and interpretation.

*Life Cycle Inventory (LCI):* The data compilation stage of LCA that quantifies all material and energy inputs and outputs — including emissions, resource consumption, and waste — for each process in the product system.

*Life Cycle Impact Assessment (LCIA):* The stage of LCA in which inventory data are translated into impact indicators such as global warming potential (GWP, in kg CO2-eq), acidification potential, human toxicity, and resource depletion, using characterization factors from methods such as ReCiPe or CML.

*Ex-ante LCA (prospective LCA):* Environmental assessment of a technology or material before it has reached industrial scale, using laboratory or pilot data, proxy technologies, and scaling relationships to estimate lifecycle impacts. Contrasts with retrospective LCA, which uses measured industrial data.

*Ensemble LCA:* Evaluation of environmental impacts across a distribution of plausible manufacturing routes, processing parameters, or background system assumptions, rather than committing to a single deterministic scenario. Produces probability distributions over impact indicators rather than point estimates, enabling uncertainty-aware decision-making.

*Carbon footprint:* The global warming potential (GWP) of a product system, expressed in kg $CO_2$-equivalent per functional unit, computed using characterization factors from the IPCC over a 100-year time horizon.

*Embodied energy:* The total primary energy consumed across all upstream processes required to produce a material or product, including extraction, processing, and manufacturing.

*Circular-by-design:* A design philosophy in which end-of-life behavior, recyclability, and environmental impact are embedded as constraints during early-stage materials or product development, rather than evaluated retrospectively.

*Life Cycle Sustainability Assessment (LCSA):* An extension of environmental LCA that integrates social LCA (S-LCA) and life cycle costing (LCC) to assess environmental, social, and economic dimensions simultaneously. The ML-LCA framework proposed in this manuscript addresses only the environmental dimension.

*Herfindahl-Hirschman Index (HHI):* A supply chain risk metric that quantifies market concentration, ranging from near zero (diversified supply) to 10,000 (single-country monopoly). Values above 2,500 indicate high supply disruption risk. HHI is not an LCA output and is outside the scope of the ML-LCA framework.

*Functional unit:* The quantified description of the performance function that a product system must deliver, serving as the reference basis for all LCA inputs and outputs. For example, "1 tonne of structural concrete with 28-day compressive strength ≥ 30 MPa."

*Product system:* The collection of unit processes linked by material and energy flows that fulfills the functional unit in an LCA study.

*Background system:* The portion of the product system that supplies inputs (electricity, heat, transport) from the broader economy and for which data are drawn from generic LCA databases such as ecoinvent or GaBi, as opposed to the foreground system, which represents processes under direct study.

**Artificial intelligence and machine learning**

*Machine learning interatomic potential (MLIP):* A computational model that approximates the quantum mechanical potential energy surface using ML regression, enabling atomistic simulations at DFT-level accuracy but orders of magnitude lower cost.

*Graph neural network (GNN):* A neural network architecture that operates directly on graph-structured data — for instance, where nodes represent atoms or material units and edges represent bonds or interactions — enabling representation learning on molecular and crystalline structures without relying on fixed-size feature vectors.

*Large language model (LLM):* A neural network with billions of parameters, pre-trained on large text corpora using self-supervised objectives, capable of generating, summarizing, and extracting structured information from text without task-specific retraining.

*Transfer learning:* An ML strategy in which a model trained on one task or domain is adapted to a related task, leveraging learned representations to reduce the data and compute required for the new domain. In materials science, this typically involves fine-tuning a model pre-trained on large compositional databases for a specific material class or property.

*Curriculum learning:* A training strategy in which examples are presented in order of increasing complexity, improving model convergence and generalization compared to random ordering.

*Surrogate model:* A computationally inexpensive approximation of a costly simulator or experiment, trained on a limited number of high-fidelity evaluations. Used for rapid screening of large candidate spaces before committing to expensive calculations or synthesis.

*Generative model:* A model that learns the distribution of training data and can sample new instances from that distribution. In materials discovery, generative models propose novel compositions or structures; examples include variational autoencoders (VAEs), generative adversarial networks (GANs), and diffusion models.

*Variational autoencoder (VAE):* A generative model that encodes input data into a continuous latent space and decodes samples from that space into new data instances. Used in materials science for inverse design by navigating the latent space toward desired property targets.

*Diffusion model:* A generative model that learns to reverse a gradual noising process, enabling generation of high-quality structured outputs such as molecular geometries or crystal structures.

*Inverse design:* The task of identifying a material composition or structure that achieves a specified set of target properties, as opposed to forward prediction, which maps structure to properties.

*Bayesian optimization (BO):* A sequential experimental design strategy that uses a probabilistic surrogate model (typically a Gaussian process) and an acquisition function to identify the next experiment most likely to improve the objective, balancing exploration of uncertain regions with exploitation of known good regions.

*Acquisition function:* A criterion used in Bayesian optimization to determine the next query point by trading off exploration and exploitation. Common choices include Expected Improvement (EI), Upper Confidence Bound (UCB), and Probability of Improvement (PI).

*Pareto frontier:* In multi-objective optimization, the set of solutions for which no objective can be improved without degrading another. Used to visualize trade-offs between performance and sustainability without imposing arbitrary weights.

*Active learning:* An ML paradigm in which the model iteratively queries the most informative new data points — selected by an acquisition strategy — to improve predictive accuracy with fewer total evaluations than random sampling.

*Reward hacking:* A failure mode in reinforcement learning or Bayesian optimization in which the agent maximizes the reward signal through unintended strategies that satisfy the metric but not the underlying goal, arising when the reward function does not fully capture the designer's intent.

*Named entity recognition (NER):* An information extraction task that identifies and classifies named entities — such as material names, properties, and synthesis conditions — in unstructured text.

*Retrosynthesis:* The process of decomposing a target molecule or material into available precursors through a sequence of feasible chemical reactions, enabling synthesis planning. Originating in organic chemistry, the concept is increasingly applied to inorganic and solid-state materials.

*Epistemic uncertainty:* Uncertainty arising from limited knowledge or data, which can in principle be reduced by collecting more information or improving models.

*Aleatoric uncertainty:* Irreducible uncertainty arising from inherent variability in a system, such as stochastic synthesis yield fluctuations, which cannot be eliminated by collecting more data.

*Scenario uncertainty:* Uncertainty arising from multiple plausible future states of the world, such as the future energy mix or regulatory environment, which cannot be resolved by additional measurement.

*Multi-fidelity data:* Data drawn from sources that differ in accuracy, cost, and resolution — for example, DFT calculations (high fidelity, expensive), classical MD simulations (intermediate fidelity), and empirical correlations (low fidelity, inexpensive) — combined within a single modeling framework to make efficient use of available information.

*Ontology:* A formal, machine-readable representation of concepts within a domain and the relationships between them, used to achieve semantic interoperability between databases and tools.

**Materials science**

*Density functional theory (DFT):* A quantum mechanical method for computing the electronic structure of materials from first principles, widely used in materials science to predict structural, mechanical, electronic, and thermodynamic properties.

*Amorphous material:* A solid lacking long-range crystalline order. Glasses are a paradigmatic example. The absence of periodicity means that standard crystal structure descriptors are inapplicable, and surrogate models must use alternative structural representations.

*SMILES (Simplified Molecular Input Line Entry System):* A notation encoding molecular structure as a linear string of characters, widely used as input representation for ML models in chemistry and polymer science.

*Polyhydroxyalkanoate (PHA):* A class of biopolyesters produced by microbial fermentation from renewable feedstocks, studied as biodegradable alternatives to petroleum-derived plastics.

*Chemically amplified resist:* A photoresist formulation used in photolithography in which a photoacid generator (PAG) produces acid upon exposure to radiation, which catalytically deprotects polymer repeat units during a post-exposure bake step to define the pattern. Widely used in EUV and DUV semiconductor lithography.

*Per- and polyfluoroalkyl substances (PFAS):* A class of synthetic chemicals containing multiple fluorine-carbon bonds, conferring exceptional chemical and thermal stability. Their environmental persistence and bioaccumulation potential are the primary regulatory concern.

*Clinker:* The calcium silicate intermediate produced by high-temperature calcination of limestone and clay in cement manufacturing, the production of which accounts for the majority of cement-related $CO_2$ emissions.

*Supplementary cementitious material (SCM):* Industrial byproducts or natural materials — such as fly ash, slag, and calcined clays — used to partially replace clinker in cement, reducing embodied $CO_2$.

*Pozzolan:* A siliceous or aluminosiliceous material that reacts with calcium hydroxide in the presence of water to form cementitious compounds. Natural pozzolans include volcanic ashes; industrial pozzolans include fly ash and silica fume.